\documentclass[
  reprint,
  superscriptaddress,
  amsmath,
  amssymb,
  aps,
  pre,
  floatfix
]{revtex4-2}

\usepackage[T1]{fontenc}
\usepackage[utf8]{inputenc}
\usepackage{graphicx}
\usepackage{bm}
\usepackage[ruled]{algorithm2e}
\usepackage{url}

\DeclareMathAlphabet{\mathcal}{OMS}{cmsy}{m}{n}

\begin{document}

\title{Evolution of cooperation with temporal information}

\author{Tianxing Zhao}
\affiliation{Center for Systems and Control, Peking University, Beijing 100871, China}

\author{Lei Zhou}
\affiliation{School of Automation, Beijing Institute of Technology, Beijing 100081, China}

\author{Aming Li}
\email{amingli@pku.edu.cn}
\affiliation{Center for Systems and Control, Peking University, Beijing 100871, China}
\affiliation{Center for Multi-Agent Research, Institute for Artificial Intelligence, Peking University, Beijing 100871, China}

\date{\today}

\begin{abstract}
Strategy learning governs the evolution of collective cooperation in multi-agent systems. Although evolutionary outcomes depend strongly on the information available to agents during strategy learning, most studies treat both the source and amount of that information as fixed over time. In reality, however, individuals are continually exposed to external information that varies over time. Here we develop a general framework for evolutionary dynamics with temporal information, which uses temporal networks to characterize dynamic changes in information available for strategy learning, with time-varying connections determining each agent’s information state. Across synthetic and empirical networks, we find that temporal information consistently promotes cooperation relative to their static counterparts. We further demonstrate that this advantage becomes more pronounced as temporal networks become sparser. This pattern arises because sparsification amplifies information heterogeneity among agents, creating unequal access to learning information that can facilitate the spread of cooperative strategies, in sharp contrast to static information formulations where heterogeneity often suppresses cooperation. Guided by this insight, we develop an interpretable algorithm that substantially enhances cooperation across diverse networks by generating learning networks with tunable heterogeneity. Our results identify temporal information as a realistic and broadly applicable mechanism for promoting collective cooperation.
\end{abstract}

\maketitle

\section{Introduction}

Understanding how cooperative behaviors emerge and prevail among self-interested individuals remains a central challenge in contemporary science~\cite{hamilton1963evolution,trivers1971evolution,nowak1992spatial,axelrod1981evolution,hauert2004snowdrift,nowak2006five,sigmund2010calculus,perc2010coevolutionary,santos2018socialnorm,hauser2019unequals}. Classical models begin with well-mixed populations, in which each individual is equally likely to interact with every other member of the group. In such populations, for canonical social dilemmas such as the prisoner's dilemma~\cite{rapoport1965prisoner}, defectors typically outperform cooperators because the latter are uniformly exposed to exploitation~\cite{hofbauer1998evolutionary,nowak2004emergence,tarnita2009set}. Real populations, however, are rarely well mixed: social, biological, and engineered systems are more accurately described as structured populations, in which interaction patterns are represented by an underlying network, whose nodes represent individuals and edges capture neighborhood relationships~\cite{watts1998collective,barabasi1999emergence,newman2003structure,chen2022synchronizability,bao2022motifs,altafini2012opinion}. In these networked systems, the architecture of local connections can foster the formation of cooperative clusters in which high-payoff cooperators are more likely than defectors to be imitated during strategy learning, thereby helping steer the system toward collective cooperation~\cite{ohtsuki_simple_2006,santos2005scalefree,taylor2007homogeneous,szabo2007graphs,santos2006heterogeneous,lieberman2005graphs,nowak2010structured,debarre2014social,allen2017Nature}.

Indeed, strategy learning governs how individuals update their behavior over time and thereby shapes which strategies and collective patterns ultimately emerge. A widely used modeling framework is imitation-based learning, in which individuals tend to adopt the strategies of more successful neighbors, that is, neighbors with higher payoffs. In such models, the set of players whose strategies and payoffs an individual can observe---the information available for learning---can strongly influence evolutionary outcomes~\cite{ohtsuki2007breaking,wang2023generalizedDB,wang2023imitation,su2022asymmetric}. For example, when individuals can access information from multiple neighbors during strategy learning, cooperation can evolve once the benefit-to-cost ratio exceeds a positive threshold. By contrast, when individuals learn from only one randomly selected neighbor, as in the so-called ``pairwise-comparison'' update rule, cooperation is disfavored irrespective of the benefit-to-cost ratio. Indeed, previous important studies have explored how the amount and composition of information available for strategy learning, as specified through different update rules, can affect the evolution of cooperation. Yet they typically treat the information state for learning as fixed over time: which neighbors can be observed and how much information is available remain unchanged during the evolutionary process.

In reality, however, individuals are continually exposed to external information that varies over time. Both the sources and the amount of available information may change as interactions unfold. Empirical studies of human social contacts have documented the intrinsically dynamic and impermanent nature of interpersonal exchanges. Examples include electronic communication through email exchanges and online question-and-answer platforms~\cite{emails_messages_data,Stack_overflow_data}, as well as face-to-face interactions in workplaces~\cite{office_data} and schools~\cite{student_data}. Such time-resolved records show that both contact partners and interaction intensities vary substantially over the course of hours and days. Consequently, learning-information configurations found to promote cooperation under static assumptions may be only shortly available in real networked systems, where structural and temporal variation are prevalent. More broadly, studies of other networked systems have shown that time-varying information-exchange patterns can substantially alter collective dynamics~\cite{scholtes2014causality,unicomb2021cascades}. This motivates a central question: how does cooperation evolve when individuals learn from time-varying information?

In this work, we investigate the evolution of cooperation in networked systems in which strategy learning is guided by temporal information. Specifically, we use temporal networks to characterize the changing information available to agents during strategy learning, with time-varying connections defining individual information states. To capture real-world contact patterns, we construct empirical temporal learning networks from recorded social-contact datasets. We show that learning with temporal information consistently promotes cooperation relative to static-information scenarios in both empirical and synthetic networks. Intriguingly, this effect is amplified in sparse networks, where information is more limited. We analytically demonstrate that this effect can be attributed to heterogeneity in information accessibility---the uneven distribution of available learning information across individuals---and that this heterogeneity is amplified by sparse temporal networks. Leveraging this insight, we develop an interpretable algorithm to generate learning networks with tunable levels of heterogeneity, further promoting the evolution of cooperation in networked systems.

\section{Model}

\begin{figure*}[!tbp]
    \centering
    \includegraphics[width=\textwidth]{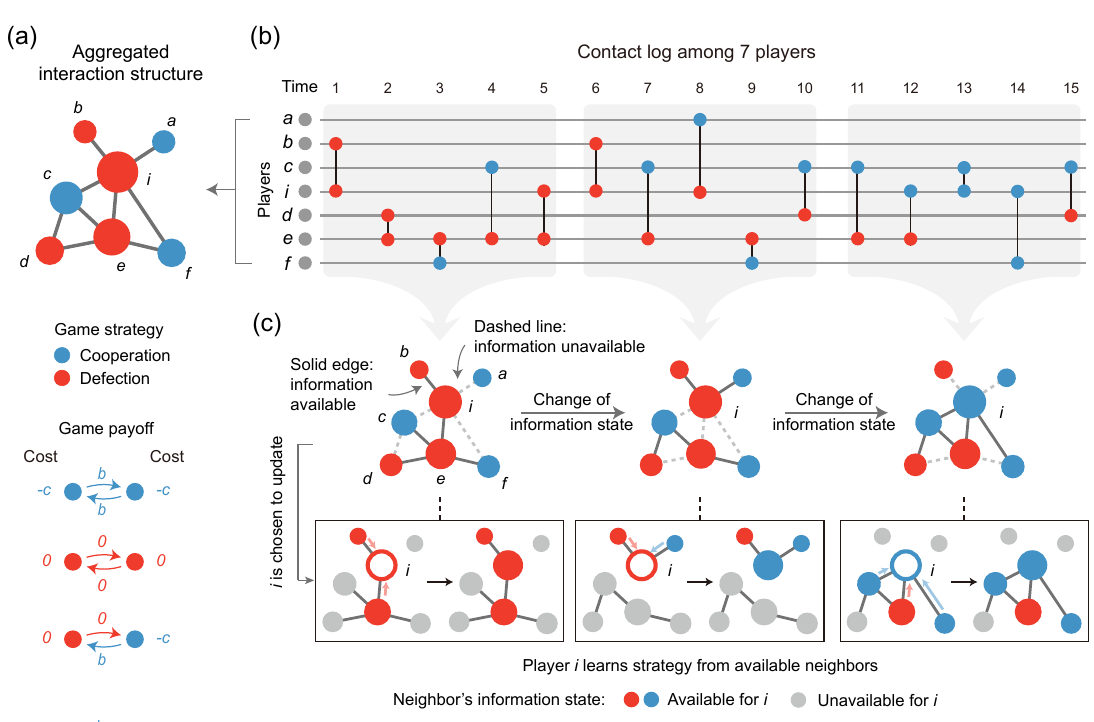}
    \caption{Illustration of the evolutionary process with temporal information.
A schematic interaction structure of seven players is illustrated in (a). The network's edges are obtained by linking any pair of players that made contact during the recorded period in (b). Each player adopts one of two strategies: cooperation (blue) or defection (red), and earns payoffs by playing the donation game with neighbors. A cooperator pays a cost $c$ to provide a benefit $b$ to its interaction partner, whereas mutual defection yields no payoff.
(b) The contact log records pairwise contacts among the same seven players over 15 time steps. Each horizontal line tracks one player, and a vertical link at time $t$ denotes a contact during the interval $(t-1,t]$. Gray shaded blocks indicate non-overlapping time windows of length $\Delta t=5$.
(c) Each time window defines a learning network by linking pairs of players that contacted at least once within that window. Solid edges indicate information that is available for strategy learning, whereas dashed edges indicate unavailable information. During an update, the chosen player $i$ learns only from its currently available neighbors, whose strategies and payoffs can be observed; unavailable neighbors are shown in gray. As the time window advances, the information state changes, providing players with new information. Starting from a single cooperator, the evolutionary process ends when cooperation takes over the population or goes extinct.}
    \label{fig:1}
\end{figure*}

We consider evolutionary game dynamics in a networked population of $N$ players. Game interactions are described by an interaction network $I$, whose nodes represent individuals and whose edges specify who plays games with whom. Strategy learning is governed by a temporal learning network $\mathcal{L}=\{L_1,\ldots,L_M\}$, represented as a sequence of snapshots that changes over the course of evolution. Each snapshot $L_m$ encodes the information available for strategy learning during a given stage of the evolutionary process: nodes represent individuals, and edges specify who can observe and imitate whom.

Each player (node) in the population adopts one of two strategies---cooperation (C) or defection (D). Cooperators pay a cost $c$ and provide a benefit $b$ to each interaction partner, whereas defectors pay no cost and provide no benefit (Fig.~\ref{fig:1}(a)). In each round, every node $i$ plays the game with its neighbors in the interaction network and receives an average payoff $f_i$. One individual is then chosen uniformly at random to update its strategy. During this update, the focal individual can access only the strategy and payoff information of its neighbors in the current learning snapshot, and imitates the strategy of neighbor $j$ with a probability proportional to $j$'s fitness. We adopt the commonly used fitness mapping $F_j = 1 + \delta f_j$, where $\delta \ge 0$ denotes the intensity of selection. After every $g$ rounds, the learning network advances to the next snapshot, thereby updating the information available for strategy learning. To facilitate comparison with prior work, we focus on the weak-selection regime ($\delta \ll 1$).

To quantify the potential for collective cooperation, we initialize the simulation by placing a single cooperator (C) at a randomly chosen node in a population of \(N\) individuals, with the remaining \(N-1\) individuals being defectors (D). The evolutionary process ends when the population reaches either full cooperation or full defection. We define the fixation probability of cooperators, \(\rho_C\), as the probability that the population eventually reaches full cooperation given this initialization. Similarly, we define \(\rho_D\) as the probability that the population reaches full defection starting from a single defector placed randomly in a population of \(N-1\) cooperators. The focus of this study is to identify the condition under which cooperation is favored over defection, namely \(\rho_C > \rho_D\).

\section{Results}
\subsection{Temporal information promotes the evolution of cooperation}

\begin{figure*}[!tbp]
    \centering
    \includegraphics[width=\textwidth]{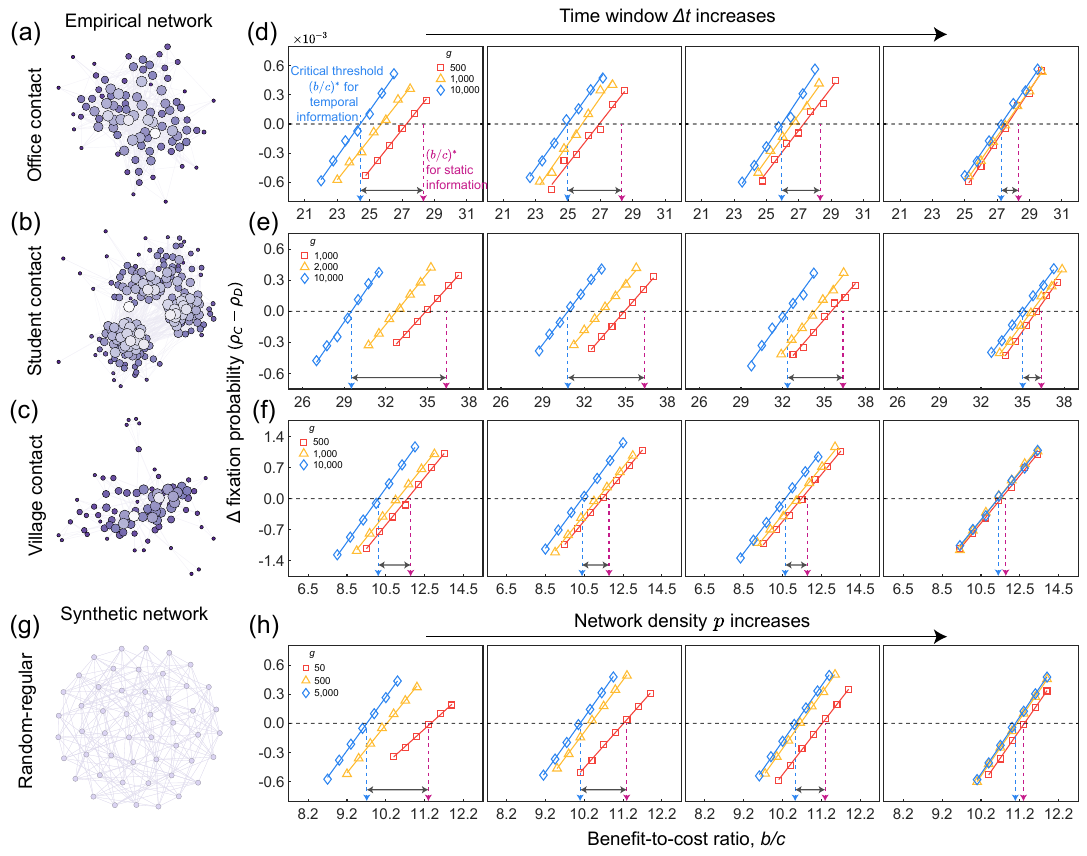}
    \caption{Temporal information promotes the fixation of cooperation in both empirical and artificial networks.
(a)--(c) Visualizations of three empirical networks capturing social relations in an office, a school and a village. (d)--(f) Evolutionary outcomes with temporal information in the three empirical networks. We plot the fixation-probability difference between cooperation and defection, $\rho_C-\rho_D$, as a function of the benefit-to-cost ratio $b/c$. Markers denote numerical simulations, and lines are linear fits. Different marker types indicate different network-switching intervals $g$. The purple vertical dashed lines mark the critical threshold $(b/c)^*$ under static information (baseline), whereas the blue vertical dashed lines mark $(b/c)^*$ with temporal information. Horizontal double-headed arrows show the reduction in $(b/c)^*$ induced by temporal information. (h) Analogous results on synthetic networks (random regular; shown in (g)), demonstrating that temporal information also lowers $(b/c)^*$ compared with the static counterpart. In (d)--(f), temporal learning networks are constructed by aggregating social contacts over non-overlapping windows of length $\Delta t$, whereas the interaction network for payoff calculation is obtained by aggregating all contacts over the recording period. We set $\Delta t=2, 6, 8, 24$ h in (d), $\Delta t=4, 6, 10, 24$ h in (e), and $\Delta t=24, 48, 72, 96$ h in (f). In (h), the interaction network is a random regular graph with $N=50$ and degree $d=8$; temporal networks are generated by uniformly sampling a fraction $p$ of edges from the static network, with $p=1/4, 3/8, 1/2, 3/4$ from left to right. We set $\delta=0.01$ throughout. Values of $(b/c)^*$ under static information are listed in Supplemental Table~1.}
    \label{fig:2}
\end{figure*}

First, we explore how learning with temporal information affects the evolution of cooperation in empirical networks. We use three empirical networks capturing social relations in an office building~\cite{office_data} (Office contact), a high school~\cite{student_data} (Student contact), and a village~\cite{village_data} (Village contact). Each dataset consists of a time-stamped log of social contacts, recording which pairs of individuals are in contact at each time. By aggregating social contacts within non-overlapping windows of length $\Delta t$, we obtain learning snapshots that encode temporal information (Fig.~\ref{fig:1}(b),(c)). Across all three empirical networks, the critical benefit-to-cost thresholds \((b/c)^*\), above which cooperation is favored over defection, are consistently lower under temporal information than under the corresponding static counterparts (Fig.~\ref{fig:2}(a)--(c); horizontal double-headed arrows mark the reduction), suggesting that temporal information relaxes the condition for cooperation to be favored. Moreover, the reduction in \((b/c)^*\) becomes larger for shorter aggregation windows \(\Delta t\), which aggregate fewer social contacts into each snapshot and correspond to sparser learning networks, indicating that cooperation is further promoted. These patterns remain unchanged when using control snapshot constructions, including randomly permuting contact timestamps and uniformly activating edges from the underlying network (Supplemental Figs.~1 and~2).

Simulations on synthetic networks yield results consistent with those on empirical networks. As in the empirical case, learning with temporal information lowers the critical threshold \((b/c)^*\) relative to the corresponding static counterpart (Fig.~\ref{fig:2}(h)). Analogous to tuning temporal information through the aggregation window \(\Delta t\) in empirical networks, we tune temporal information in synthetic networks by controlling the density of learning snapshots. Specifically, in each snapshot, we independently and uniformly activate a fraction \(p\) of edges from the underlying network to form the learning network. We find that smaller \(p\) yields lower \((b/c)^*\), consistent with the empirical results. These qualitative findings are further supported across various classes of synthetic networks (Supplemental Fig.~3).

We further validate the robustness of these results to network-switching speeds and update rules. We use $\Delta(b/c)^*$ to quantify the change in the critical threshold relative to the static baseline, where $\Delta(b/c)^*<0$ indicates a lower threshold and thus facilitated cooperation. In previous temporal network models, in which both the interaction and learning networks are time-varying, frequent network switching can suppress cooperation ($\Delta(b/c)^*>0$) or even make cooperation consistently disfavored (Supplemental Fig.~4). By contrast, in our temporal information framework, cooperation is preserved ($\Delta(b/c)^*\approx 0$) or enhanced ($\Delta(b/c)^*<0$) even when the learning network switches frequently (i.e., at small \(g\), such as \(g=10\); Supplemental Fig.~5(a)). The facilitation of cooperation strengthens as the network-switching interval \(g\) increases. For a given \(g\), sparser learning networks, which contain fewer connections, typically produce a larger reduction in \((b/c)^*\), consistent with the trends observed when varying \(\Delta t\) or \(p\). These conclusions are robust to population sizes and mean degrees (Supplemental Fig.~5(b),(c)). A similar pattern emerges under the IM update rule, in which individuals can not only learn from their current neighbors but also retain the option to keep their strategy. Temporal information again lowers $(b/c)^*$, and the facilitation is stronger for sparser learning networks (Supplemental Fig.~5(d)--(f)).

\subsection{The role of information heterogeneity}

\begin{figure*}[!tbp]
    \centering
    \includegraphics[width=\textwidth]{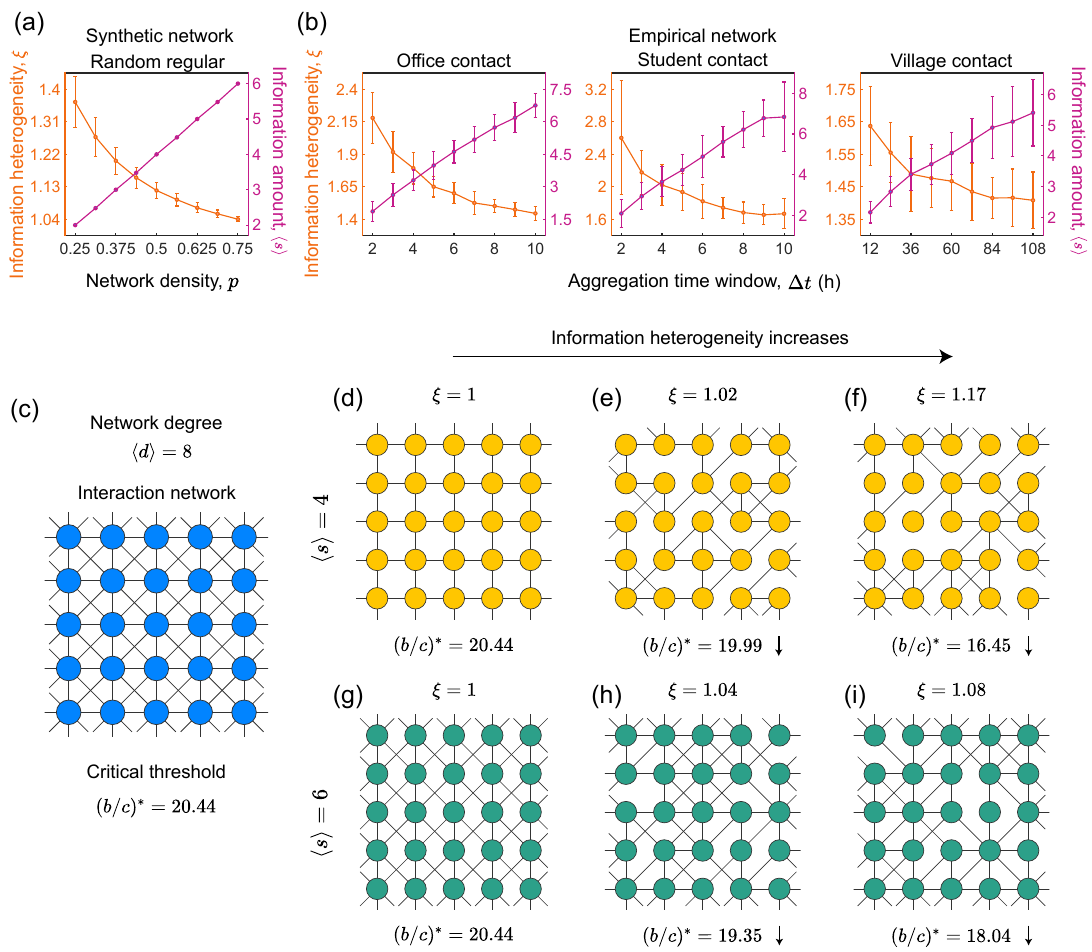}
    \caption{Information heterogeneity promotes cooperation.  (a),(b) Network density and aggregation time windows shape the information state of the learning network, characterized by the average amount of information \(\langle s\rangle\) and the information heterogeneity \(\xi\). Error bars denote standard deviations across temporal snapshots. To isolate how these two features of the information state influence cooperation, we fix the interaction network and vary only the learning network. (c) A lattice with degree \(\langle d\rangle = 8\) serves as the shared interaction network. (d)--(i) Learning networks are constructed as subgraphs of (c) with prescribed average information \(\langle s\rangle\) (top row, \(\langle s\rangle = 4\); bottom row, \(\langle s\rangle = 6\)) and increasing heterogeneity \(\xi\) from left to right (values above panels). The corresponding critical thresholds \((b/c)^*\) are reported below each panel. Using (c) as both interaction and learning networks yields \((b/c)^* = 20.44\). When heterogeneity is held fixed at $\xi=1$ (homogeneous information access), changing the mean information \(\langle s\rangle\) from \(8\) ((c)) to \(4\) ((d)) or \(6\) ((g)) leaves \((b/c)^*\) unchanged (\(20.44\)), indicating that \(\langle s\rangle\) alone does not affect the threshold when information heterogeneity is absent. In contrast, increasing \(\xi\) at fixed \(\langle s\rangle\) ((d)--(f) and (g)--(i)) monotonically lowers \((b/c)^*\). Across all six learning networks, the critical threshold is inversely related to \(\xi\).}
   \label{fig:3}
\end{figure*}

Having demonstrated that temporal information promotes collective cooperation, we next seek to understand the underlying mechanism. The effect of time-varying learning information on evolutionary outcomes arises from changes in the structure of the learning network, which defines the information state of the system. This structure shapes learning information in two ways: it determines the amount of information each individual can access, and the distribution of information across individuals. We quantify these features using the average amount of available information, \(\langle s\rangle\), and information heterogeneity, \(\xi\), which captures how unevenly learning information is distributed across individuals (see Supplementary Information).

We start by examining how different implementations of temporal information affect the information states of the evolutionary system. Across empirical and synthetic systems, the implementation of temporal information is controlled by the aggregation window \(\Delta t\) and the network density \(p\), respectively. Varying either control parameter produces consistent changes in learning information (Fig.~\ref{fig:3}(a),(b)): decreasing either parameter, a change associated in both cases with lower thresholds for cooperation, reduces the average amount of available information, \(\langle s\rangle\), while increasing information heterogeneity, \(\xi\). This suggests that sparser learning networks reduce the average amount of information available to individuals while amplifying disparities in information access across individuals. From a network perspective, the former corresponds to a reduction in the average degree of the learning network \(L\), whereas the latter corresponds to an increase in the degree heterogeneity of \(L\), as each agent's information access is determined by its learning neighbors.

A natural question is whether the reduction in average degree is the main driver of the observed promotion of cooperation. This intuition is consistent with previous insight from network reciprocity, where lower degree can facilitate cooperation~\cite{ohtsuki_simple_2006}, as captured for example by the simple rule \(b/c>\langle k\rangle\). To answer this question, we construct regular subgraphs of a lattice with different average degrees. This allows us to tune the degree of the learning network while preserving degree regularity. However, we find that lowering the average degree of the learning network \(L\) alone does not affect the critical threshold for cooperation when heterogeneity is absent (\(\xi=1\); Fig.~\ref{fig:3}(d),(g)).

We next turn to degree heterogeneity, the other structural change induced by sparsification. Although degree heterogeneity is ubiquitous in real systems, it has often been associated with weaker support for the fixation of cooperation. Indeed, previous studies of fixation probability have demonstrated that degree heterogeneity can impede the evolution of cooperation~\cite{allen2017Nature,fotouhi2019cooperation}. For instance, among regular, Erd\H{o}s--R\'enyi and scale-free networks with fixed population size $N$ and average degree, the threshold for cooperative success increases as degree heterogeneity increases~\cite{meng2024dynamics}. These canonical findings typically assume that the learning and interaction structures are identical (i.e., $L=I$).

Surprisingly, our analyses reveal that information heterogeneity, captured by the degree heterogeneity of the learning network \(L\), can substantially promote cooperation. To isolate this effect, we construct learning networks with the same average amount of available information but different levels of information heterogeneity, while keeping the interaction structure fixed (Fig.~\ref{fig:3}(c)--(i)). This comparison separates the effect of information heterogeneity from that of information scarcity. The resulting critical thresholds show that, with the average amount of information held fixed, the threshold for cooperation decreases as information heterogeneity increases. These findings point to information heterogeneity---the uneven distribution of available learning information across individuals---as a key mechanism by which temporal information promotes the evolution of cooperation. This mechanism further explains why sparser learning networks provide stronger support for cooperation, as sparsification amplifies information heterogeneity rather than merely reducing the average amount of available information; in the absence of such heterogeneity, evolutionary outcomes remain unchanged across different amounts of available information (Supplemental Fig.~6).

\subsection{Theoretical analyses}
\begin{figure}[!tbp]
    \centering
    \includegraphics[width=\columnwidth]{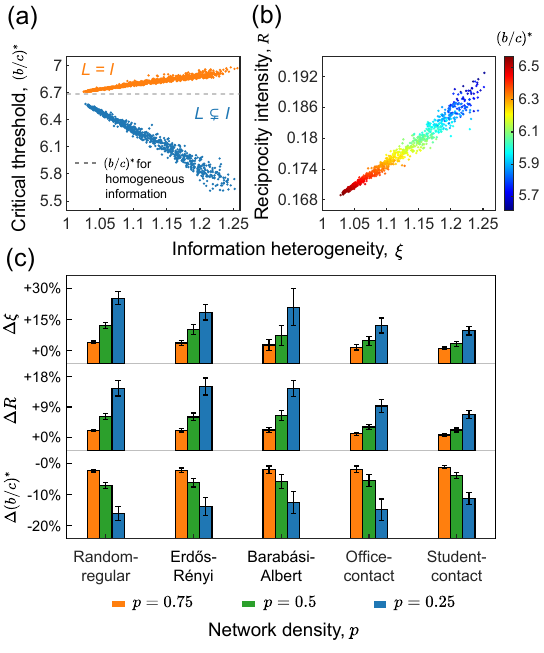}
    \caption{Quantitative relationship between information heterogeneity and evolutionary outcome.
(a) Critical benefit-to-cost threshold $(b/c)^*$ versus information heterogeneity $\xi$. The dashed line marks the homogeneous-information baseline, $\xi=1$ and $(b/c)^*=6.68$. When learning and interaction networks coincide ($L=I$; upper branch), increasing $\xi$ raises the threshold; when the learning network is independently structered as a connected subgraph of the interaction network ($L\subsetneq I$; lower branch), increasing $\xi$ instead markedly lowers it.
(b) For the evolutionary systems in the lower branch of (a), reciprocity intensity $\mathcal{R}$ increases with $\xi$, while the corresponding $(b/c)^*$ decreases, as indicated by color. Each point represents one system. This joint variation is consistent with a reciprocity-mediated mechanism by which information heterogeneity promotes cooperation.
(c) Generality across synthetic and empirical networks. For each base network and sampling fraction \(p\in\{0.75,0.5,0.25\}\), \(1{,}000\) learning networks are generated by activating a fraction \(p\) of interaction edges, giving \(3{,}000\) networks per base network. Percentage changes in $\xi$, $\mathcal{R}$ and $(b/c)^*$ are measured relative to the \(p=1\) baseline ($L=I$). Lower \(p\) consistently increases $\xi$ and $\mathcal{R}$ while decreasing $(b/c)^*$. Bars show means \(\pm\) s.d. In (a),(b), \(N=100\), \(\langle d\rangle=6\), and subgraph learning networks are generated by activating different fractions of edges of the regular interaction network. In (c), RRG and ER have \(N=100\) and \(\langle d\rangle=8\); BA has \(N=100\), \(\langle d\rangle=8\) and \(\gamma=3\).}
    \label{fig:4}
\end{figure}
In addition to the comparative experiments showing that heterogeneous learning networks support cooperation more strongly than homogeneous counterparts, we further provide a theoretical explanation for the relationship between information heterogeneity and the critical threshold through the lens of network reciprocity.

We first give a closed-form expression for the critical benefit-to-cost threshold $(b/c)^*$ above which cooperation is favored for a population with interaction network $I$ and learning network $L$ (see Supplementary Information)
\begin{equation}
\left( \frac{b}{c} \right)^* =
\frac{\sum_{i,j=1}^N \pi_i p_{ij}^{(2,0)} \eta_{ij}}{
\sum_{i,j=1}^N \pi_i p_{ij}^{(2,1)} \eta_{ij} -
\sum_{i,j=1}^N \pi_i p_{ij}^{(0,1)} \eta_{ij}}.
\label{exact_bcr}
\end{equation}
Here $p_{ij}^{(n,m)}$ is the transition probability from $i$ to $j$ for an $(n,m)$-step walk, with $n$ steps on $L$ followed by $m$ steps on $I$, and \(\eta_{ij}\) is the expected coalescence time between nodes $i$ and $j$ under a discrete-time coalescing random walk on the learning network $L$.

Using Eq.~(\ref{exact_bcr}), we compute the critical threshold $(b/c)^*$ for a suite of systems that share the same interaction network but have different learning networks spanning a broad range of information heterogeneity (Fig.~\ref{fig:4}(a), lower region). A pronounced inverse correlation emerges between information heterogeneity $\xi$ and the critical threshold \((b/c)^*\), in contrast to the positive correlation under the conventional setting $L=I$ (Fig.~\ref{fig:4}(a), upper region). This reversal indicates that information heterogeneity---equivalently, degree heterogeneity of $L$---favors the evolution of cooperation when $L \subsetneq I$.

To intuitively understand how information heterogeneity fosters cooperation, we derive a more interpretable expression for \((b/c)^{*}\) using a mean-field approximation and the recurrence relations for coalescence times
\begin{equation}
\left( \frac{b}{c} \right)^*\approx
\frac{\hat{\eta} - 1}{
\left( \hat{\eta} + 1 \right)
\mathcal{R} - 2}.
\label{bcr_approx}
\end{equation}
The quantity $\hat{\eta}$ is the average value of $\eta_{ij}$ over all pairs $(i,j)$ with $i\neq j$. Here, \(\mathcal{R}=\sum_{i,j=1}^N \frac{\pi_i}{2}\left(p_{ij}q_{ji}+q_{ij}p_{ji}\right)\) is called the reciprocity intensity, where \(\pi_i\) is the reproductive value of individual \(i\), proportional to its degree in the learning network, \(p_{ij}\) denotes the one-step transition probability from node \(i\) to node \(j\) on the learning network, and \(q_{ij}\) denotes the corresponding transition probability on the interaction network. In intuitive terms, \(\mathcal{R}\) measures how strongly payoff interactions are aligned with strategy learning, so that individuals receiving cooperative benefits are more likely to copy the donor's strategy. Such local feedback facilitates the spread and persistence of cooperation among interacting partners, and Eq.~(\ref{bcr_approx}) shows that larger \(\mathcal{R}\) lowers the critical threshold.

We further derive an analytical expression for the reciprocity intensity $\mathcal{R}$, highlighting the role of information heterogeneity, in terms of summary statistics of the interaction and learning networks
\begin{equation}
\begin{aligned}
\mathcal{R}
=\frac{1}{2}\Bigg[&\frac{1}{\langle d\rangle}
+\frac{1}{\langle d\rangle}
\Big(\xi+\operatorname{Cov}_{\mathrm{E}}\!\big(k_i,\tfrac{1}{k_j}\big)\Big)
\\
&+\frac{1}{\langle k\rangle^2}
\operatorname{Cov}\!\big(k^2,\tfrac{1}{d}\big)
+\frac{1}{\langle k\rangle}
\operatorname{Cov}\!\big(k,\tfrac{1}{d}\big)
-\eta_I\Bigg].
\end{aligned}
\label{P11_exact}
\end{equation}
Here, $\langle d\rangle=\sum_{i=1}^N d_i/N$ and $\langle k\rangle=\sum_{i=1}^N k_i/N$ denote the average degrees of the interaction network $I$ and the learning network $L$, respectively (with $d_i$ and $k_i$ the corresponding node degrees). Note that $k_i$ can also be interpreted as the amount of information available to node $i$. The degree heterogeneity of $L$ is quantified by $\xi=\langle k^{2}\rangle/\langle k\rangle^{2}$, which is equivalent to our measure of information heterogeneity. $\operatorname{Cov}_{\mathrm{E}}(\cdot,\cdot)$ denotes covariance computed over the set of directed edges induced by $L$, whereas $\operatorname{Cov}(\cdot,\cdot)$ denotes covariance across nodes. The term $\eta_I$ is a correction that depends on degree heterogeneity in the interaction network $I$ and vanishes for degree-homogeneous interaction networks (see Supplementary Information).

With this analytical tool, a deeper explanation of the advantage conferred by information heterogeneity emerges through the lens of network reciprocity. For the systems with heterogeneous learning networks considered in the lower region of Fig.~\ref{fig:4}(a), we plot their reciprocity intensity $\mathcal{R}$ against information heterogeneity $\xi$ in Fig.~\ref{fig:4}(b), with each point representing one system and color indicating the corresponding critical threshold $(b/c)^*$. We find that $\mathcal{R}$ increases monotonically with $\xi$. This relationship is captured by our analytical expression for $\mathcal{R}$ in Eq.~\eqref{P11_exact}, which shows how increasing $\xi$ amplifies reciprocity. The color gradient further indicates that larger $\xi$, and hence larger $\mathcal{R}$, is systematically associated with a lower $(b/c)^*$. Together, these results make the mechanism explicit: increasing information heterogeneity strengthens network reciprocity, which in turn lowers the critical threshold for cooperation.

By activating different fractions of edges across five interaction networks---random regular (RRG), Barabási--Albert (BA), Erdős--Rényi (ER), and the empirical Office and Student contact networks---we provide a unifying account of how increasing information heterogeneity promotes collective cooperation. For each interaction network, we generate a broad set of learning networks with different network densities, and compute the changes in information heterogeneity \(\Delta\xi\), reciprocity intensity \(\Delta \mathcal{R}\), and critical threshold \(\Delta(b/c)^*\) relative to the baseline with identical learning and interaction networks (Fig.~\ref{fig:4}(c)). Across the five populations, we observe that sparser learning networks increases $\xi$, which coincides with the increases in reciprocity intensity $\mathcal{R}$ and reductions in $(b/c)^*$, thereby bridging our theoretical analyses with previous findings on empirical and synthetic networks (Fig.~\ref{fig:2}).

Our theory can also be applied to dynamics governed by the IM update rule (Supplementary Note~4). As under DB updating, heterogeneity in $L$ continues to promote cooperation under IM updating. Moreover, the average degree of $L$ plays an important role under the IM rule: a higher average degree favors cooperation by lowering the relative weight of personal information in strategy learning (Supplemental Fig.~12).

\subsection{Promoting cooperation by tuning information heterogeneity}
\begin{figure*}[!tbp]
    \centering
    \includegraphics[width=0.99\textwidth]{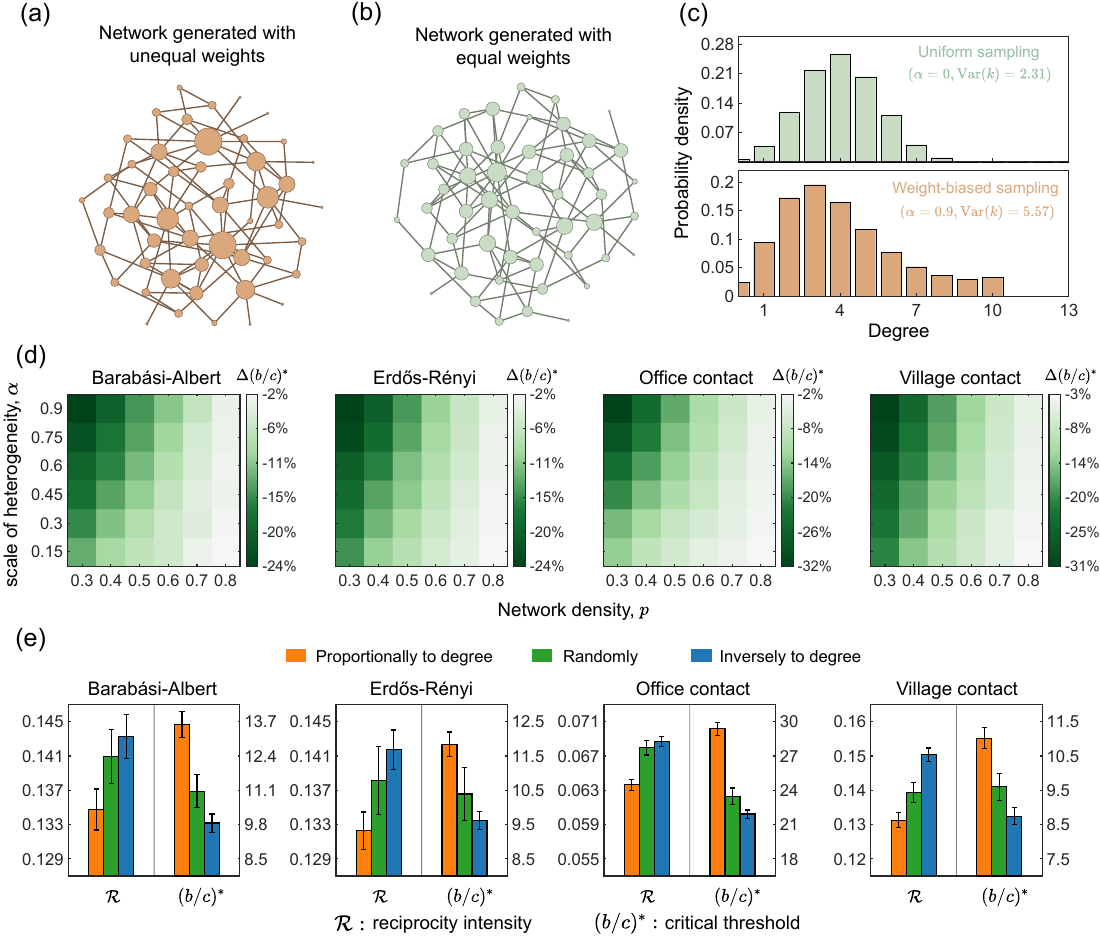}
    \caption{Effectiveness and controllability of GenUINe. (a),(b) Visualizations of networks generated using GenUINe (left) and uniform edge sampling (right). (c) Degree distributions for networks generated by uniform edge sampling (equal weights \(\alpha=0\)) and networks generated by GenUINe (unequal weights \(\alpha=0.9\), as in (a)). (d) Heat maps showing how GenUINe promotes cooperation across four bases: Barabási--Albert (BA), Erdős--Rényi (ER), and the empirical Office contact and Village contact networks. In each panel, the base network serves as the interaction network, while GenUINe generates learning networks controlled by the scale of heterogeneity \(\alpha\) and the network density \(p\). Colors encode the relative change in the critical threshold, $\Delta(b/c)^*$, with darker colors indicating larger threshold reductions; the gradient shows that increasing $\alpha$ and decreasing $p$ progressively lower $(b/c)^*$. (e) Influence of node-weighting schemes adopted in GenUINe when generating learning networks, evaluated on the same four heterogeneous bases. Bars report the reciprocity intensity \(\mathcal{R}\) and the critical threshold \((b/c)^*\) under three policies: inverse-degree weighting (blue), random weights (green), and proportional-to-degree weighting (orange). Across all bases, inverse-degree weighting yields the highest \(\mathcal{R}\) and the lowest \((b/c)^*\), random weighting is intermediate, and proportional-to-degree weighting produces the lowest \(\mathcal{R}\) and the highest \((b/c)^*\). The base network used in (a)--(c) is regular with \(N=50\) and \(\langle d\rangle=10\), and the generated subnetworks have mean degree \(\langle k\rangle=4\).
In (d),(e), BA and ER have \(N=50\) and \(\langle d \rangle=8\) (BA with \(\gamma=3\)); in (e), \(p=0.5\) and \(\alpha=0.9\), with \(100\) learning networks generated per setting.}
    \label{fig:5}
\end{figure*}

As a step toward designing cooperation-promoting learning environments, can our analytical findings be translated into an general design principle? To address this question, we develop Generate Unequal Information Networks (GenUINe), an interpretable generative algorithm for constructing learning networks with tunable information heterogeneity. The basic idea is to assign unequal sampling weights to nodes according to a prescribed ranking, and then use these weights to guide the formation of learning connections from the underlying network. When the scale of heterogeneity \(\alpha=0\), all nodes are sampled uniformly; as \(\alpha\) increases, sampling weights become increasingly uneven, making high-weight nodes more likely to acquire learning connections. In this way, GenUINe generates learning networks with a prescribed density while systematically tuning the heterogeneity of information access. The node ranking further determines how this heterogeneity is aligned with the underlying network structure. The full procedure is given in Algorithm~\ref{Algorithm1}.

GenUINe reliably generates networks with tunable heterogeneity. Even when the input network $G$ is degree-homogeneous, the algorithm can generates subnetworks with power-law-like degree profiles (Fig.~\ref{fig:5}(a),(c)). Relative to uniform sampling (\(\alpha=0\)), introducing unequal node weights markedly increases degree heterogeneity. We then apply GenUINe to two synthetic networks (BA and ER) and two empirical networks (Office contact and Village contact), varying two control parameters: the heterogeneity scale \(\alpha\) and the network density \(p\). The four heatmaps in Fig.~\ref{fig:5}(d) reveal a consistent pattern across all networks: increasing \(\alpha\) or decreasing \(p\) monotonically lowers the critical threshold \((b/c)^*\). Taken together, Fig.~\ref{fig:5}(a)--(d) establish both the effectiveness and the controllability of GenUINe.

% \begingroup
% \makeatletter\@twocolumnfalse\makeatother
\begin{algorithm}[!t]
\footnotesize
\SetAlgoLined
\DontPrintSemicolon
\KwIn{Base network \(G=(V,E)\) with \(|V|=N\),\newline
scale of heterogeneity \( \alpha \in [0, 1) \),\newline
network density \(p\in(0,1]\).}
\KwOut{A subgraph $H$ on \(V\) with\newline
\(|E(H)|=\lfloor p\,|E|\rfloor\).}

\textbf{Step 1: Node Initialization} \;
Draw a permutation \(\pi\) of the node set \(V\) (random by default).\;
Let \(r_i\in\{1,\dots,N\}\) be the rank (position) of node \(i\) in \(\pi\) (\(r_i=1\) is first).\;

\For{$i\gets1$ \KwTo $N$}{
    Assign node \(i\) a weight \( p_i = r_{i}^{-\alpha} \).\;
}
Initialize $H$ as $N$ isolated nodes.\;

\textbf{Step 2: Node Sampling and Edge Formation} \;
\While{$|E(H)|<\lfloor p\cdot|E(G)|\rfloor$}{
    Independently sample two distinct nodes \(i,j\) with probabilities:
    \[
    \begin{aligned}
    P(i) &= \frac{p_i}{\sum_{k}p_k},\\
    P(j) &= \frac{p_j}{\sum_{k}p_k}.
    \end{aligned}
    \]
    \If{\((i,j)\notin E(H)\) \textbf{and} \((i,j)\in E(G)\)}{
        Add edge \((i,j)\) to \(H\).\;
    }
}

\Return \(H\)\;

\caption{\textbf{Gen}erate \textbf{U}nequal \textbf{I}nformation \textbf{Ne}tworks (GenUINe)}
\label{Algorithm1}
\end{algorithm}
% \endgroup

\begin{figure*}[!tbp]
    \centering
    \includegraphics[width=\textwidth]{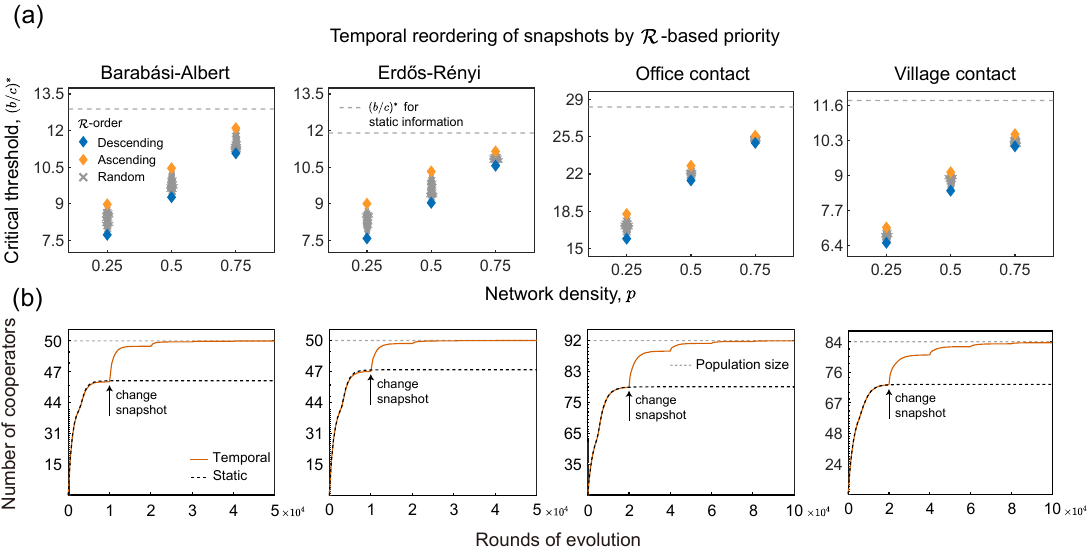}
    \caption{Temporal reordering of learning networks further promotes the fixation of cooperation. (a) Effects of temporal reordering of heterogeneous learning snapshots by snapshot-level reciprocity \(\mathcal{R}\) on the fixation of cooperation. For each base network (BA, ER, Office contact, Village contact) and network density \(p\in\{0.25,0.5,0.75\}\), we first generate a set of 10 GenUINe snapshots ($\alpha=0.9$, inverse-degree weighting) and then assemble three temporal sequences: \(\mathcal{R}\)-descending (blue diamonds), random order (gray crosses), and \(\mathcal{R}\)-ascending (orange diamonds). Markers report the resulting critical thresholds \((b/c)^*\); the horizontal gray dashed line marks the static-information baseline. Ordering snapshots by decreasing \(\mathcal{R}\) yields the lowest \((b/c)^*\), random ordering is intermediate, and \(\mathcal{R}\)-ascending ordering gives the highest \((b/c)^*\), demonstrating that the temporal arrangement of information can be used to further promote cooperation. (b) Time course of the number of cooperators for systems that use either a single snapshot as a static learning network (black dashed lines) or a sequence of snapshots as a temporal learning network (orange lines). Each jump coincides with a snapshot change, showing how temporal rewiring reconnects otherwise disjoint components and provides a spatio-temporally connected pathway for cooperation to reach fixation, whereas the static snapshot stalls at a partial level. The network-switching interval is \(g=1000\) in (a) for BA and ER, and \(g=2000\) in (a) for Office and Village. In (b), the network densities are \(p=0.5, 0.5, 0.25\) and \(0.5\) from left to right. For \((b/c)^*\) with static information under different settings, see Supplemental Table~1.}
    \label{fig:6}
\end{figure*}

We next examine how the node-level allocation of heterogeneous information access affects evolutionary outcomes. Using the same four base networks, we compare three node-weighting policies for generating heterogeneous learning networks: inverse-degree weighting, in which high-degree nodes are assigned smaller sampling weights by being placed later in the ranking; random weighting, as used in the heatmaps in Fig.~\ref{fig:5}(d); and degree-proportional weighting, in which high-degree nodes are assigned higher sampling weights. Across all base networks, inverse-degree weighting yields the highest reciprocity intensity \(\mathcal{R}\) and the lowest \((b/c)^*\), whereas degree-proportional weighting yields the lowest \(\mathcal{R}\) and the highest \((b/c)^*\); random weighting lies between these two cases (Fig.~\ref{fig:5}(e)). This suggests that cooperation is most strongly supported when heterogeneous information access is negatively aligned with node degree, thereby strengthening reciprocity and lowering the threshold for cooperation. This pattern corroborates our theoretical analysis of \(\mathcal{R}\), as Eq.~(3) shows that the covariance terms \(\operatorname{Cov}(k^2,1/d)\) and \(\operatorname{Cov}(k,1/d)\) enter with positive coefficients. Accordingly, configurations in which degrees in \(L\) are negatively correlated with degrees in \(I\)---equivalently, positively correlated with \(1/d\)---increase \(\mathcal{R}\). Inverse-degree weighting produces precisely this anticorrelation, whereas degree-proportional weighting induces the opposite tendency and thereby reduces \(\mathcal{R}\).

Extending these principles to temporal information, we next ask how to exploit the temporal degree of freedom---the order in which heterogeneous learning networks are presented---to further lower the critical threshold. Since the analysis above links information heterogeneity to reciprocity intensity \(\mathcal{R}\), and \(\mathcal{R}\) in turn tracks reductions in \((b/c)^*\), we use \(\mathcal{R}\) as a data-driven priority signal for scheduling snapshots. For each base network, we generate a set of GenUINe snapshots under a high-heterogeneity regime with inverse-degree weighting and a fixed sampling fraction \(p\), and vary only their temporal order to construct temporal learning networks. We compare three schedules in which high-\(\mathcal{R}\) snapshots appear early, randomly, or late in the sequence, corresponding to \(\mathcal{R}\)-descending, random, and \(\mathcal{R}\)-ascending orders, respectively. Across all systems, the \(\mathcal{R}\)-descending schedule consistently yields the lowest critical threshold \((b/c)^*\), the random schedule yields intermediate values, and the \(\mathcal{R}\)-ascending schedule performs worst (Fig.~\ref{fig:6}(a)). The effect is strongest when sparse, highly heterogeneous snapshots are ordered in decreasing $\mathcal{R}$, reducing the critical threshold $(b/c)^*$ by nearly half.

Fig.~\ref{fig:6}(b) illustrates another fundamental advantage of temporal information for the fixation of cooperation. Cooperators spread only along connected paths, yet strong degree heterogeneity can fragment an individual snapshot, as hubs or communities monopolize links while peripheral nodes remain disconnected. This trade-off is reflected by the dashed lines in Fig.~\ref{fig:6}(b): when cooperation spreads on a single learning snapshot, it reaches only part of the population, primarily the nodes in the largest connected component. By contrast, when heterogeneous snapshots are presented sequentially, each snapshot change rewires these components, generating stepwise increases in the number of cooperators and eventually enabling fixation. In this sense, pronounced information heterogeneity makes cooperation more likely to emerge, whereas network temporality supplies the connected routes through which the cooperative strategy can ultimately reach the entire population.

\section{Discussion}

We have shown that, relative to static information, strategy learning with temporal information consistently promotes cooperation across both empirical and synthetic networks, and that this advantage becomes stronger in sparser learning networks, where available information is more limited. This pattern differs from earlier results for static incomplete information under death–birth dynamics, where limiting available information by random sampling—without restructuring the learning network—was found not to alter the condition for cooperative success. Our analysis attributes the distinctive effect of temporal information to the increased heterogeneity of learning networks, which strengthens network reciprocity and lowers the critical threshold for cooperation to be favored. This mechanism, in turn, motivates GenUINe as an interpretable framework for designing learning networks that further enhance cooperation.

At the microscopic level, reciprocating an interaction partner through strategy imitation helps seed cooperative clusters. Reciprocity arises when a cooperative act is followed by strategy spread in the same direction, which in turn generates a cooperative return. A substantial body of influential work has established the central role of network reciprocity in the evolution of cooperation~\cite{allen2017Nature,su2022asymmetric,meng2025promoting}, yet a general theoretical framework for explaining and harnessing this effect remains lacking. Our analysis of reciprocity intensity provides a quantitative account of how network structure shapes pairwise reciprocity. In particular, increasing heterogeneity in the learning network substantially strengthens reciprocity. This effect follows from the way reciprocity intensity couples payoff interactions with learning transitions: individuals with greater learning access carry more reproductive weight in strategy transmission, whereas those with more limited access assign greater weight to each available learning source. When these two roles meet along the same interaction, their asymmetry strengthens reciprocal learning, making cooperative benefits more likely to feed back into imitation and subsequent strategy spread. Guided by these insights, GenUINe operationalizes a reciprocity-based design principle for promoting cooperation, and can even restore the support for cooperation in populations whose original interaction networks do not favor its evolution by generating temporal learning networks that strengthen reciprocity.

From a modeling perspective, our approach differs from most previous studies of evolutionary dynamics on temporal networks, which typically do not distinguish between interaction and learning networks~\cite{li2020temp_evolution,su2023strategy,cardillo2014evolutionary,kun2009evol_dynamical_graph}. Such formulations are natural extensions of static network models and have yielded important insights into the role of time-varying structure in evolutionary dynamics. For the present question, however, they make it difficult to examine the role of temporal information as it is entangled with structural variation that can itself favor cooperation~\cite{ohtsuki_simple_2006}. By separating temporal learning from interaction structure, we identify temporal information as a distinct mechanism and show that it alone can generate substantial cooperative gains. In this view, temporal learning networks constitute a distinct control layer: by reshaping who learns from whom, and crucially, in what temporal order (Fig.~\ref{fig:6}(a)), they can significantly promote the evolution of cooperation. By linking this informational layer to the interpretable quantity of reciprocity intensity, our study opens a broader perspective on how collective behavior can be understood and shaped through the temporal architecture of environmental information.

% \section*{Data availability}

% All data supporting the findings of this study are included in this paper and its Supplementary Information. The empirical temporal network datasets analysed in this study are publicly available through the SocioPatterns collaboration, as cited in the manuscript.

% \section*{Code availability}
% The code used in this study was written in MathWorks MATLAB R2022a, Python 3.8.3 and Julia 1.10, and is available at \url{https://github.com/HenryZhao625/Temporal-information}.

% \section*{Competing interests}
% The authors declare no competing interests.


\begin{thebibliography}{99}

\bibitem{hamilton1963evolution}
Hamilton, W.~D.
The evolution of altruistic behavior.
\emph{Am. Nat.} \textbf{97}, 354--356 (1963).

\bibitem{trivers1971evolution}
Trivers, R.~L.
The evolution of reciprocal altruism.
\emph{Q. Rev. Biol.} \textbf{46}, 35--57 (1971).

\bibitem{nowak1992spatial}
Nowak, M.~A. \& May, R.~M.
Evolutionary games and spatial chaos.
\emph{Nature} \textbf{359}, 826--829 (1992).

\bibitem{axelrod1981evolution}
Axelrod, R. \& Hamilton, W.~D.
The evolution of cooperation.
\emph{Science} \textbf{211}, 1390--1396 (1981).

\bibitem{hauert2004snowdrift}
Hauert, C. \& Doebeli, M.
Spatial structure often inhibits the evolution of cooperation in the snowdrift game.
\emph{Nature} \textbf{428}, 643--646 (2004).

\bibitem{nowak2006five}
Nowak, M.~A.
Five rules for the evolution of cooperation.
\emph{Science} \textbf{314}, 1560--1563 (2006).

\bibitem{sigmund2010calculus}
Sigmund, K.
\emph{The Calculus of Selfishness}
(Princeton Univ. Press, Princeton, 2010).


\bibitem{perc2010coevolutionary}
Perc, M. \& Szolnoki, A.
Coevolutionary games---a mini review.
\emph{BioSystems} \textbf{99}, 109--125 (2010).

\bibitem{santos2018socialnorm}
Santos, F.~P., Santos, F.~C. \& Pacheco, J.~M.
Social norm complexity and past reputations in the evolution of cooperation.
\emph{Nature} \textbf{555}, 242--245 (2018).

\bibitem{hauser2019unequals}
Hauser, O.~P., Hilbe, C., Chatterjee, K. \& Nowak, M.~A.
Social dilemmas among unequals.
\emph{Nature} \textbf{572}, 524--527 (2019).


\bibitem{rapoport1965prisoner}
Rapoport, A., Chammah, A.~M. \& Orwant, C.~J.
\emph{Prisoner's Dilemma: A Study in Conflict and Cooperation}
(Univ. of Michigan Press, Ann Arbor, 1965).

\bibitem{hofbauer1998evolutionary}
Hofbauer, J. \& Sigmund, K.
\emph{Evolutionary Games and Population Dynamics}
(Cambridge Univ. Press, Cambridge, 1998).

\bibitem{nowak2004emergence}
Nowak, M.~A., Sasaki, A., Taylor, C. \& Fudenberg, D.
Emergence of cooperation and evolutionary stability in finite populations.
\emph{Nature} \textbf{428}, 646--650 (2004).

\bibitem{tarnita2009set}
Tarnita, C.~E., Antal, T., Ohtsuki, H. \& Nowak, M.~A.
Evolutionary dynamics in set structured populations.
\emph{Proc. Natl. Acad. Sci. U.S.A.} \textbf{106}, 8601--8604 (2009).

\bibitem{watts1998collective}
Watts, D.~J. \& Strogatz, S.~H.
Collective dynamics of `small-world' networks.
\emph{Nature} \textbf{393}, 440--442 (1998).

\bibitem{barabasi1999emergence}
Barab{\'a}si, A.-L. \& Albert, R.
Emergence of scaling in random networks.
\emph{Science} \textbf{286}, 509--512 (1999).

\bibitem{newman2003structure}
Newman, M.~E.~J.
The structure and function of complex networks.
\emph{SIAM Rev.} \textbf{45}, 167--256 (2003).

\bibitem{chen2022synchronizability}
Chen, G.
Searching for best network topologies with optimal synchronizability: a brief review.
\emph{IEEE/CAA J. Autom. Sin.} \textbf{9}, 573--577 (2022).


\bibitem{bao2022motifs}
Bao, X., Hu, Q., Ji, P., Lin, W., Kurths, J. \& Nagler, J.
Impact of basic network motifs on the collective response to perturbations.
\emph{Nat. Commun.} \textbf{13}, 5301 (2022).

\bibitem{altafini2012opinion}
Altafini, C.
Dynamics of opinion forming in structurally balanced social networks.
\emph{PLoS ONE} \textbf{7}, e38135 (2012).

\bibitem{ohtsuki_simple_2006}
Ohtsuki, H., Hauert, C., Lieberman, E. \& Nowak, M.~A.
A simple rule for the evolution of cooperation on graphs and social networks.
\emph{Nature} \textbf{441}, 502--505 (2006).

\bibitem{santos2005scalefree}
Santos, F.~C. \& Pacheco, J.~M.
Scale-free networks provide a unifying framework for the emergence of cooperation.
\emph{Phys. Rev. Lett.} \textbf{95}, 098104 (2005).

\bibitem{taylor2007homogeneous}
Taylor, P.~D., Day, T. \& Wild, G.
Evolution of cooperation in a finite homogeneous graph.
\emph{Nature} \textbf{447}, 469--472 (2007).

\bibitem{szabo2007graphs}
Szab\'o, G. \& F\'ath, G.
Evolutionary games on graphs.
\emph{Phys. Rep.} \textbf{446}, 97--216 (2007).

\bibitem{santos2006heterogeneous}
Santos, F.~C., Pacheco, J.~M. \& Lenaerts, T.
Evolutionary dynamics of social dilemmas in structured heterogeneous populations.
\emph{Proc. Natl. Acad. Sci. U.S.A.} \textbf{103}, 3490--3494 (2006).

\bibitem{lieberman2005graphs}
Lieberman, E., Hauert, C. \& Nowak, M.~A.
Evolutionary dynamics on graphs.
\emph{Nature} \textbf{433}, 312--316 (2005).

\bibitem{nowak2010structured}
Nowak, M.~A., Tarnita, C.~E. \& Antal, T.
Evolutionary dynamics in structured populations.
\emph{Phil. Trans. R. Soc. B} \textbf{365}, 19--30 (2010).

\bibitem{debarre2014social}
D\'ebarre, F., Hauert, C. \& Doebeli, M.
Social evolution in structured populations.
\emph{Nat. Commun.} \textbf{5}, 3409 (2014).

\bibitem{allen2017Nature}
Allen, B., Lippner, G., Chen, Y.-T., Fotouhi, B., Momeni, N., Yau, S.-T. \& Nowak, M.~A.
Evolutionary dynamics on any population structure.
\emph{Nature} \textbf{544}, 227--230 (2017).

\bibitem{ohtsuki2007breaking}
Ohtsuki, H., Nowak, M.~A. \& Pacheco, J.~M.
Breaking the symmetry between interaction and replacement in evolutionary dynamics on graphs.
\emph{Phys. Rev. Lett.} \textbf{98}, 108106 (2007).


\bibitem{wang2023generalizedDB}
Wang, C. \& Szolnoki, A.
Evolution of cooperation under a generalized death-birth process.
\emph{Phys. Rev. E} \textbf{107}, 024303 (2023).

\bibitem{wang2023imitation}
Wang, X., Zhou, L., McAvoy, A. \& Li, A.
Imitation dynamics on networks with incomplete information.
\emph{Nat. Commun.} \textbf{14}, 7453 (2023).


\bibitem{su2022asymmetric}
Su, Q., Allen, B. \& Plotkin, J.~B.
Evolution of cooperation with asymmetric social interactions.
\emph{Proc. Natl. Acad. Sci. U.S.A.} \textbf{119}, e2113468118 (2022).


\bibitem{emails_messages_data}
Holme, P. \& Saram\"aki, J.
Temporal networks.
\emph{Phys. Rep.} \textbf{519}, 97--125 (2012).

\bibitem{Stack_overflow_data}
Paranjape, A., Benson, A.~R. \& Leskovec, J.
Motifs in temporal networks.
In \emph{Proceedings of the Tenth ACM International Conference on Web Search and Data Mining}
(eds Tomkins, A. \& Zhang, M.) 601--610
(Association for Computing Machinery, New York, 2017).

\bibitem{office_data}
G\'enois, M., Vestergaard, C.~L., Fournet, J., Panisson, A., Bonmarin, I. \& Barrat, A.
Data on face-to-face contacts in an office building suggest a low-cost vaccination strategy based on community linkers.
\emph{Netw. Sci.} \textbf{3}, 326--347 (2015).

\bibitem{student_data}
Fournet, J. \& Barrat, A.
Contact patterns among high school students.
\emph{PLoS ONE} \textbf{9}, e107878 (2014).


\bibitem{scholtes2014causality}
Scholtes, I., Wider, N., Pfitzner, R., Garas, A., Tessone, C.~J. \& Schweitzer, F.
Causality-driven slow-down and speed-up of diffusion in non-Markovian temporal networks.
\emph{Nat. Commun.} \textbf{5}, 5024 (2014).


\bibitem{unicomb2021cascades}
Unicomb, S., I{\~n}iguez, G., Gleeson, J.~P. \& Karsai, M.
Dynamics of cascades on burstiness-controlled temporal networks.
\emph{Nat. Commun.} \textbf{12}, 133 (2021).

\bibitem{village_data}
Ozella, L., Paolotti, D., Lichand, G., Rodr\'iguez, J.~P., Haenni, S., Phuka, J., Leal-Neto, O.~B. \& Cattuto, C.
Using wearable proximity sensors to characterize social contact patterns in a village of rural Malawi.
\emph{EPJ Data Sci.} \textbf{10}, 46 (2021).

\bibitem{fotouhi2019cooperation}
Fotouhi, B., Momeni, N., Allen, B. \& Nowak, M.~A.
Evolution of cooperation on large networks with community structure.
\emph{J. R. Soc. Interface} \textbf{16}, 20180677 (2019).

\bibitem{meng2024dynamics}
Meng, Y., Cornelius, S.~P., Liu, Y.-Y. \& Li, A.
Dynamics of collective cooperation under personalised strategy updates.
\emph{Nat. Commun.} \textbf{15}, 3125 (2024).

\bibitem{meng2025promoting}
Meng, Y., McAvoy, A. \& Li, A.
Promoting collective cooperation through temporal interactions.
\emph{Proc. Natl. Acad. Sci. U.S.A.} \textbf{122}, e2509575122 (2025).

\bibitem{li2020temp_evolution}
Li, A., Zhou, L., Su, Q., Cornelius, S.~P., Liu, Y.-Y., Wang, L. \& Levin, S.~A.
Evolution of cooperation on temporal networks.
\emph{Nat. Commun.} \textbf{11}, 2259 (2020).

\bibitem{su2023strategy}
Su, Q., McAvoy, A. \& Plotkin, J.~B.
Strategy evolution on dynamic networks.
\emph{Nat. Comput. Sci.} \textbf{3}, 763--776 (2023).

\bibitem{cardillo2014evolutionary}
Cardillo, A., Petri, G., Nicosia, V., Sinatra, R., G\'omez-Garde\~nes, J. \& Latora, V.
Evolutionary dynamics of time-resolved social interactions.
\emph{Phys. Rev. E} \textbf{90}, 052825 (2014).

\bibitem{kun2009evol_dynamical_graph}
Kun, {\'A}. \& Scheuring, I.
Evolution of cooperation on dynamical graphs.
\emph{BioSystems} \textbf{96}, 65--68 (2009).

\end{thebibliography}
\end{document}